# Quasinormal-mode analysis of grating spectra at fixed incidence angles

Alexandre Gras,[1] Wei Yan,[1,⊥] Philippe Lalanne[1,*]
Affiliations :
[1]LP2N, Institut d'Optique Graduate School, CNRS, Univ. Bordeaux, 33400 Talence, France
[⊥]Present address at Westlake University, Hangzhou, China

*Corresponding author: philippe.lalanne@institutoptique.fr

**Abstract.** Grating spectra exhibit sharp variations of the scattered light, known as grating anomalies. The latter are due to resonances that have fascinated specialists of optics and physics for decades and are nowadays used in many applications. We present a comprehensive theory of grating anomalies, and develop a formalism to expand the field scattered by metallic or dielectric gratings into the basis of its natural resonances, thereby enabling the possibility to reconstruct grating spectra measured for fixed illumination angles as a sum over every individual resonance contribution with closed-form expressions. This gives physical insights into the spectral properties and a direct access to the resonances to engineer the spectral response of gratings and their sensitivity to tiny perturbations.

The efficiencies of gratings as a function of the wavelength may present peaks, dips or anomalies generated by the excitation of leaky photonic of plasmonic modes. Yet another example is found in Fig. 1. This is well known since U. Fano introduced a surface- plasmon model to analyze light diffraction by metallic gratings and explained Wood's anomalies [1]. Nowadays, grating resonances have many applications for biosensing, photodetectors, photovoltaics, light emission, optical processing, metamaterials … and their theoretical analysis for harnessing light-matter interaction remains of great importance.

The theory of grating anomalies was pioneered by a milestone work by Hessel and Oliner [2] and was then followed by a series of works summarized in Refs. [3–5], which contributed to the systematic development of a phenomenological study of grating anomalies through the poles and zeros of the scattering operator, the so-called "pology". Poles were indifferently computed by considering a real frequency ω (equal to the driving laser frequency) and looking for complex in-plane wave-vectors $\tilde{k}(\omega)$ or angles of incidence $\sin(\tilde{\theta}(\omega))$, or by considering a fixed angle of incidence $\theta$, and looking for complex frequencies $\tilde{\omega}(\theta)$. Great insight in the physics of grating anomalies was achieved by tracking the pole trajectories in the complex plane as some parameters, e.g. the grating depth, are tuned [4]. The frequency poles $\tilde{\omega}$, i.e. the natural resonances, have a profound meaning (these poles correspond to the quasinormal-modes or QNMs hereafter). They define important quantities such as the resonance frequency, $\text{Re}(\tilde{\omega})$, or the inverse of the mode lifetime, $2\text{Im}(\tilde{\omega})$. The theory of grating anomalies changed little during several decennia, and the pology has been used to analyze or engineer various anomalous grating effects [6]. In 2014, Vial and his colleagues published a paper that contains many important results [7]. Assuming one-dimensional gratings made of nondispersive materials, they computed many QNMs (poles) at complex frequencies, then normalized the QNMs and computed their excitation coefficients, to finally reconstruct the field scattered by an incident plane wave in the QNM basis. The formalism was subsequently used to design spectral filters in the infrared [8]. The new possibilities offered by the availability of stable methods to normalize QNMs were also exploited for deriving closed-form expressions of the changes of grating resonance frequency and linewidth due to tiny refractive index changes [9,10]. Recently, several numerical methods to compute and normalize the QNMs of plasmonic nanoresonators, including metal gratings and plasmonic crystals, were successfully benchmarked, establishing standards for the computation and normalization of QNMs of dispersive resonators [11]. These initiatives are part of broader, more comprehensive studies on QNM-expansion formalisms for analyzing light interaction with resonances of open non-Hermitian electromagnetic systems [12], which are presently knowing rapid progresses [13].

In all recent works [7,8,9,11], QNMs are computed for a fixed in-plane Bloch-wavevector $\boldsymbol{k}_p$. We will refer to those modes as $\boldsymbol{k}_p$-QNMs. Here we rather consider $\boldsymbol{\eta}$-QNMs, which are computed for a fixed angle of incidence, i.e. a fixed "directionality" vector $\boldsymbol{\eta} = \boldsymbol{k}_p c/\omega$, a real vector that depends on the incident plane wave and the refractive index $n_i$ of the incident medium, but not on the frequency. Specifically, consider a grating that is periodic along the $x$ and $y$ directions, and a typical scattering problem that the grating is illuminated by a plane wave with a fixed angle of incidence having $\boldsymbol{k}_p = \frac{\omega n_i}{c}[\sin(\theta)\cos(\phi)\hat{\boldsymbol{x}} + \sin(\theta)\sin(\phi)\hat{\boldsymbol{y}}]$, where $\theta$ and $\phi$ are the angular components of the spherical coordinates that specify the incidence angles in a medium with refractive index $n_i$ (assumed to be non-dispersive). We see that $\boldsymbol{k}_p$ changes with frequency, while $\boldsymbol{\eta} = n_i(\sin(\theta)\cos(\phi)\hat{\boldsymbol{x}} + \sin(\theta)\sin(\phi)\hat{\boldsymbol{y}})$ is independent of frequency.

We emphasize that $\boldsymbol{k}_p$-QNMs and $\boldsymbol{\eta}$-QNMs are different and have different normalizations; they have different field distributions, frequencies . We also stress that grating experiments are only indirectly related to band-structure computations and that $\boldsymbol{\eta}$-QNMs are the resonances that are effectively revealed by any experimental grating spectra measured for fixed incident angles [4]. Similarly, in a sensing experiment, the shift in resonance frequency is due to a perturbation of $\boldsymbol{\eta}$-QNMs, not $\boldsymbol{k}_p$-QNMs, except if one deliberately records the spectra by simultaneously tuning the angle of incidence for every frequency measurement. Hereafter we propose a method to compute $\boldsymbol{\eta}$-QNMs using the COMSOL Multiphysics environment. By extending a recent theoretical framework based on the auxiliary-field method [13] to gratings, we introduce a new norm and orthogonality relation, see Eq. (3), which explicitly depends on $\boldsymbol{\eta}$, and establish a closed-form expression, see Eq. (4), for the excitation coefficients of $\boldsymbol{\eta}$-QNMs. We finally test the formalism by proposing the first reconstruction of the spectrum of a metallic grating in a QNM basis and by establishing a new perturbation theory of gratings.

Hereafter, the tilde notation will be used to indicate (often complex) quantities related to QNMs and the time convention $\exp(-i\omega t)$ is adopted. All materials are assumed to be isotropic, non-magnetic, and reciprocal, for simplicity. We further assume that the grating material is dispersive and that its permittivity follows a single pole Lorentz model, $\varepsilon_m(\omega) = \varepsilon_\infty - \varepsilon_\infty \omega_p^2 (\omega^2 - \omega_0^2 + i\omega\gamma)^{-1}$. We denote the QNM fields by $[\widetilde{\mathbf{H}}_\eta(\mathbf{r}), \widetilde{\mathbf{E}}_\eta(\mathbf{r})] = [\widecheck{\mathbf{h}}_\eta(\mathbf{r}), \widetilde{\mathbf{e}}_\eta(\mathbf{r})] \exp(i\frac{\omega}{c}\boldsymbol{\eta} \cdot \mathbf{r})$, $\widecheck{\mathbf{h}}_\eta(\mathbf{r})$ and $\widetilde{\mathbf{e}}_\eta(\mathbf{r})$ being periodic functions with the same periodicity as the grating. The following is largely inspired from a recent work on a rigorous modal analysis of plasmonic nanoresonators, which is extended here to encompass periodicity in two spatial directions. We avoid lengthy mathematical derivations hereafter to focus on the main results and refer the readers to details in [13]. For dispersive materials, QNMs are source-free solutions of a nonlinear eigenproblem [12]. We use auxiliary-fields [13–15] to linearize Maxwell's equations, and introduce two auxiliary fields in relation with the Lorentz model for the permittivity, the polarization $\widetilde{\mathbf{P}}_\eta = \varepsilon_\infty \widetilde{\omega}_p^2 (\widetilde{\omega}^2 - \omega_0^2 + i\widetilde{\omega}\gamma)^{-1} \widetilde{\mathbf{E}}_\eta$ and the current $\widetilde{\mathbf{J}}_\eta = -i\widetilde{\omega}\widetilde{\mathbf{P}}_\eta$, at the QNM frequency $\widetilde{\omega}$. Using the constitutive relation $\widetilde{\mathbf{D}}_\eta = \varepsilon_m(\widetilde{\omega})\widetilde{\mathbf{E}}_\eta$ and reformulating Maxwell's equations with the augmented electromagnetic vector $[\widetilde{\mathbf{H}}_\eta, \widetilde{\mathbf{E}}_\eta, \widetilde{\mathbf{P}}_\eta, \widetilde{\mathbf{J}}_\eta] = [\widecheck{\mathbf{h}}_\eta, \widetilde{\mathbf{e}}_\eta, \widetilde{\mathbf{p}}_\eta, \widetilde{\mathbf{j}}_\eta] \exp(i\frac{\omega}{c}\boldsymbol{\eta} \cdot \mathbf{r})$, we obtain a generalized linear eigenvalue equation defining the $\boldsymbol{\eta}$-QNMs:

$$H\widetilde{\boldsymbol{\Psi}}_\eta = \widetilde{\omega} M_\eta \widetilde{\boldsymbol{\Psi}}_\eta, \tag{1}$$

with $\widetilde{\boldsymbol{\Psi}}_\eta = [\widecheck{\mathbf{h}}_\eta(\mathbf{r}), \widetilde{\mathbf{e}}_\eta(\mathbf{r}), \widetilde{\mathbf{p}}_\eta(\mathbf{r}), \widetilde{\mathbf{j}}_\eta(\mathbf{r})]^T$ (the superscript "T" denotes the transpose operation) and

$$H = \begin{bmatrix} 0 & -i\mu_0^{-1}\nabla\times & 0 & 0 \\ i\varepsilon_\infty^{-1}\nabla\times & 0 & 0 & -i\varepsilon_\infty^{-1} \\ 0 & 0 & 0 & i \\ 0 & i\omega_p^2\varepsilon_\infty & -i\omega_0^2 & -i\gamma \end{bmatrix} \tag{2a}$$

$$M_\eta = \begin{bmatrix} 1 & -\frac{\mu_0^{-1}}{c}\boldsymbol{\eta}\times & 0 & 0 \\ \frac{\varepsilon_\infty^{-1}}{c}\boldsymbol{\eta}\times & 1 & 0 & 0 \\ 0 & 0 & 1 & 0 \\ 0 & 0 & 0 & 1 \end{bmatrix}. \tag{2b}$$

It is noticeable that the operators $H$ and $M_\eta$ are frequency-independent, so that Eq. (1) defines a generalized linear eigenproblem that is easily solved with classical routines after discretization with finite elements for instance [13].

The derivation of the normalization and orthogonality relations for $\boldsymbol{\eta}$-QNMs in the the augmented space follows the approach in [13]. The sole difference is that, the derivation in Ref. [13] was originally developed for non-periodic dispersive resonators for which $\boldsymbol{\eta} = 0$ and $M_\eta$ is an identity matrix, while the in-plane periodicity here imposes two QNMs with opposite $\boldsymbol{\eta}$'s [16,11], $\widetilde{\boldsymbol{\Psi}}_\eta \exp(i\frac{\omega}{c}\boldsymbol{\eta}\cdot\mathbf{r})$ and $\widetilde{\boldsymbol{\Psi}}_{-\eta} \exp(-i\frac{\omega}{c}\boldsymbol{\eta}\cdot\mathbf{r})$ which always exist in pairs if the grating constitutive materials are reciprocal [16], as we assume here. The new $\boldsymbol{\eta}$-QNMs' biorthonormality relation then reads as

$$\langle \widetilde{\boldsymbol{\Psi}}^*_{-\eta,n} | DM_\eta | \widetilde{\boldsymbol{\Psi}}_{\eta,m} \rangle_{V_{cell}} \equiv \iiint_{V_{cell}} \boldsymbol{\Psi}^T_{-\eta,n}(r) DM_\eta \boldsymbol{\Psi}_{\eta,m}(r) d^3r = \delta_{mn}, \tag{3}$$

where we introduce the subscripts $m$ and $n$ to label two different QNMs with opposite $\eta'$s, the diagonal matrix $\boldsymbol{D} = \text{diag}\left[-\mu_0, \varepsilon_\infty, \frac{\omega_0^2}{\varepsilon_\infty \omega_p^2}, -\frac{1}{\varepsilon_\infty \omega_p^2}\right]$ and the Kronecker delta $\delta_{mn} = 1$ if $m = n$ and otherwise 0. The integral in Eq. (3) is performed over the volume of the grating unit cell, $V_{cell}$, which is delimited by its periodic boundaries and the outermost boundaries of the perfectly-matched layers (PMLs) [17]. For gratings with a mirror symmetry plane, like the one of Fig. 1, note that $\boldsymbol{\Psi}_{-\eta}$ can be straightforwardly deduced from $\boldsymbol{\Psi}_\eta$[11], without any computation. Otherwise, two QNMs have to be computed independently, one with $\eta$ and the other with $-\eta$. The normalization condition of Eq. (3) for $m = n$, can be rewritten as $\iiint_{V_{cell}} \frac{\partial \widetilde{\omega}\varepsilon(\widetilde{\omega})}{\partial \widetilde{\omega}} \tilde{\mathbf{e}}_{-\eta} \cdot \tilde{\mathbf{e}}_\eta - \mu_0 \tilde{\mathbf{h}}_{-\eta} \cdot \tilde{\mathbf{h}}_\eta - \frac{1}{c}\eta \cdot (\tilde{\mathbf{h}}_{-\eta} \times \tilde{\mathbf{e}}_\eta + \tilde{\mathbf{e}}_{-\eta} \times \tilde{\mathbf{h}}_\eta) d^3 r = 1$, which depends on $\eta$ and is different from those used in earlier works [7,8,9,11]. We additionally note that it is also possible to normalize the $\eta$-QNMs without PMLs, for instance by using the RCWA and by adapting the general method proposed in [18].

Let us now consider the reconstruction problem of the scattered field upon illumination by an incident plane wave $[\mathbf{H}_{inc}, \mathbf{E}_{inc}] = [\mathbf{h}_{inc}, \mathbf{e}_{inc}] \exp(i\frac{\omega}{c}\eta \cdot \mathbf{r})$ with the "directionality" vector $\eta$ and the (real) frequency $\omega$. In the scattered-field formalism used for the expansion, the scattered field is defined by a local change $\Delta\varepsilon(\omega, \mathbf{r})$ ($\Delta\varepsilon \neq 0$ for $\mathbf{r} \in V_{res}$) for a background permittivity $\varepsilon_b$ so that $\varepsilon_b + \Delta\varepsilon$ is equal to the permittivity distribution of the grating geometry. Similarly, the background field $[\mathbf{h}_b, \mathbf{e}_b, 0,0] \exp(i\frac{\omega}{c}\eta \cdot \mathbf{r})$ is defined as the electromagnetic field that is solution of Maxwell's equations for the background permittivity distribution $\varepsilon_b(\omega, \mathbf{r})$ upon illumination by the incident plane wave, see Annex 2 in [12]. From the orthonormality condition, it is straightforward [13] to derive closed-form expressions for the excitation coefficients $\alpha_m$ of the QNM expansion of the scattered field $\boldsymbol{\Psi}_{sca}(\omega, \eta) = \sum_{m=1}^{M} \alpha_m(\omega) \widetilde{\boldsymbol{\Psi}}_{\eta,m}$

$$\alpha_m = \frac{\widetilde{\omega}_m}{\omega - \widetilde{\omega}_m} \langle \tilde{\mathbf{e}}^*_{-\eta,m} | \varepsilon(\widetilde{\omega}_m) - \varepsilon_b | \mathbf{e}_b \rangle_{V_{res}} + \langle \tilde{\mathbf{e}}^*_{-\eta,m} | \varepsilon_b - \varepsilon_\infty | \mathbf{e}_b \rangle_{V_{res}}. \tag{4}$$

The bra-ket product in Eq. (4) is defined by $\langle \tilde{\mathbf{e}}^*_{-\eta,m} | f(\mathbf{r}) | \mathbf{e}_b \rangle_{V_{res}} = \iiint_{V_{res}} f(\mathbf{r}) \tilde{\mathbf{e}}_{-\eta,m} \cdot \mathbf{e}_b d^3\mathbf{r}$. It represents an overlap integral involving the background field, the $-\eta$-QNM electric field, and some weighting function $f(\mathbf{r})$ that is nonnull over the finite volume $V_{res}$. In relation with Fig. 1c, $V_{res}$ corresponds to the grooves, $\varepsilon_b(\omega)$ is the permittivity distribution of an air-metal interface at $y = 0$ and $\Delta\varepsilon = \varepsilon_g - \varepsilon_m(\omega)$ with $\varepsilon_g = 1$ the permittivity of the air grooves. In summary, referring to Eq. (4), we have $\varepsilon(\widetilde{\omega}) = \varepsilon_g$, $\varepsilon_\infty = 1$, and $\varepsilon_b = \varepsilon_m(\omega)$. The total field $\boldsymbol{\Psi}_{tot}$ then reads as the sum of the scattered field and the incident field, i.e., $\boldsymbol{\Psi}_{tot} \exp(i\frac{\omega}{c}\eta \cdot \mathbf{r}) = (\boldsymbol{\Psi}_{sca} + [\mathbf{h}_b, \mathbf{e}_b, 0,0]) \exp(i\frac{\omega}{c}\eta \cdot \mathbf{r})$.

To demonstrate the effectiveness of the method, we consider a gold grating that efficiently absorbs light over the visible (we have optimized the groove depth for that purpose) and near infrared spectral range. The grating, whose characteristics are described in the caption of Fig. 1, is composed of tiny rectangular air grooves, see the inset in Fig. 1a, and is illuminated by a plane wave impinging from air with an angle of incidence $\theta = 30°$ and polarized with a magnetic-field parallel to the grooves (TM polarization). The black dots show the specular reflectance computed with the Rigorous Coupled Wave Analysis (RCWA) [19,20]. Since 401 Fourier harmonics are retained in the computation, the computed data are highly accurate and be used as reference data hereafter. Using COMSOL Multiphysics, we build a model of the grating unit cell. The periodicity along the $x$-direction is enforced with periodic boundary conditions, and the outgoing-wave condition is fulfilled in the $y$-direction using PMLs. We then compute the eigenmodes with the built-in eigensolver of COMSOL using the augmented field formulation of Eq. (1). The computational time per eigenmode is a few seconds with a standard desktop computer. We then normalized the QNMs using Eq. (3). The magnetic-fields of the 7 dominant QNMs, labelled A, B, … G, which dictate the positions and widths of the 7 main resonance dips, are displayed in the upper panel of Fig. 1a.

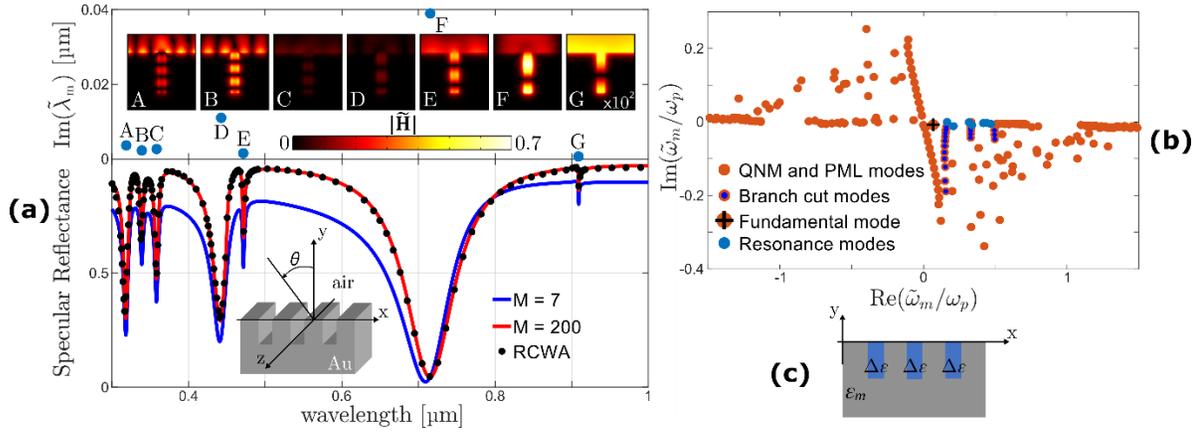

Figure 1 : (a) Reconstruction of the reflectance spectrum of a gold grating composed of tiny rectangular air grooves (see the inset) illuminated by a plane wave with an angle of incidence $\theta = 30°$ in the $x - y$ plane and TM-polarized. The reconstructed specular reflectance spectra with $M = 7$ and 200 modes retained in the expansion are shown with the blue and red curves, respectively, and are compared with reference data obtained with the RCWA (black dots). The magnetic-field moduli of the $M = 7$ dominant normalized QNMs, labelled A, B, … G, dictate the positions and widths of the 7 main resonance dips. They are computed with COMSOL Multiphysics and are shown in the upper panel. The grating period is 600 nm, and the groove depth and width are 350 nm and 60 nm, respectively. All computations are performed by assuming a Drude-Lorentz permittivity $\varepsilon_m(\omega) = 1 - \omega_p^2(\omega^2 - \omega_0^2 + i\omega\gamma)^{-1}$ for gold, with $\omega_p = 1.26\ 10^{16}\ rad.s^{-1}$, $\omega_0 = 0$, and $\gamma = 0.0112\omega_p$. (b) Complex plane representation of sample of the modes used in the expansion for $M = 200$. (c) For the scattered-field formulation, cf. Eq. (4), we consider that the scattering object is a periodic array of air holes drilled into a metal substrate, thereby implying that a local change of the background permittivity $\Delta\varepsilon = 1 - \varepsilon_m$.

We further reconstruct the scattered field in the QNM basis using the formula of Eq. (4), and then derive the specular reflectance from the Rayleigh expansion. First we only consider the seven dominant QNMs in the expansion ($m = 1, … 7$). The specular reflection computed for $M = 7$ is shown with the blue curve. Qualitative agreement with the RCWA data is achieved, particularly for resonance features. We then sort the computed QNMs according to their decreasing impact on the reconstructed specular reflection spectra [13] (the impact is known analytically, hence sorting is straightforward) and reconstruct the scattered field with the first $M = 200$ most impactful ones, see the red curve. Now a quantitative agreement is achieved. Thanks to computational results obtained with the real frequency solver of COMSOL Multiphysics, we have additionally verified that the small residual difference with the RCWA data is primarily due to a numerical dispersion between the Fourier and finite-element methods, rather than to the specific choice of the truncation rank $M = 200$. A mode-by-mode study of the convergence of the QNM expansion shows that the reconstruction accuracy is particularly impacted by QNMs that lie outside the frequency window of interest, especially the one in the infrared with a single antinode inside the groove (see Fig. 1b), and by a myriad of purely numerical eigenmodes, referred to as PML modes in the literature [7,12,13] that are due to the branch cuts in the complex plane imposed by the passing of diffracted orders at some frequencies, around which these modes aggregate (see Fig. 1b) [7]. In contrast with the case of localized resonators [13], we do not foresee the present method as competing with more traditional computational methods, e.g. the RCWA, but we emphasize that the reconstruction with 7 QNMs qualitatively reproduces all the salient features (the peaks, dips and Fano shape) of the spectrum and thus comprehensively restores the physics at works, which could be a precious added value for design. Advanced details concerning the numerical implementation of the method can be found at the webpage of the authors group in the QNMEig workpackage freeware. These include a released COMSOL model sheet and the companion Matlab script used to compute the $\alpha_m$'s and reconstruct the scattered field.

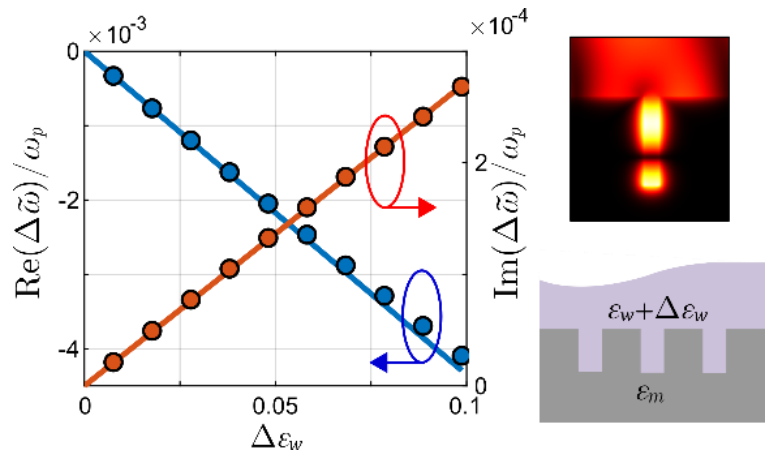

Figure 1 : Resonance changes, $\text{Re}(\Delta\widetilde{\omega})$ and $\text{Im}(\Delta\widetilde{\omega})$, induced by a relative permittivity change $\Delta\varepsilon_w$ of the permittivity (lower-right panel). The gold grating is the same as in Fig. 1 except that we consider that it is immersed in a medium of relative permittivity $\varepsilon_w$. Left main panel: fully numerical calculations and analytical predictions are represented by circles and solid curves respectively. Right top: magnetic-field distribution of the unperturbed QNM, $\widetilde{\omega} = (0.1606 - 0.0093i)\omega_p$.

An important application of grating resonances is refractive index biosensing [21], in which concentration changes of relevant substances in solutions are detected by measurements of changes in the resonance spectrum. Important metrics are the relative change in resonance frequency and linewidth induced by a tiny permittivity perturbation $\Delta\varepsilon_w(\mathbf{r})$ housed within the volume $V_{pert}$. At the design stage, the spectrum change can be predicted by repeated computations, scanning the frequency response of the perturbed and unperturbed systems, this requires repeated computations for every perturbation instance, so it is more convenient to consider cavity perturbation theory.

Hereafter, the same approach as in [22] is adopted. The $\boldsymbol{\eta}$-QNM of the unperturbed system with a permittivity $\varepsilon(\mathbf{r},\widetilde{\omega})$ is slightly modified due to the permittivity perturbation. We denote by $(\mathbf{e}'_\eta,\widetilde{\omega}')$ the QNM of the perturbed system with its new resonant frequency $\widetilde{\omega}' = \widetilde{\omega} + \Delta\omega$. By applying the divergence theorem to $(\mathbf{e}_\eta,\mathbf{h}_\eta,\widetilde{\omega})$ and $(\mathbf{e}'_{-\eta},\mathbf{h}'_{-\eta},\widetilde{\omega}')$ and performing a first order Taylor expansion of the permittivity, $\varepsilon'(\mathbf{r},\omega') = \varepsilon(\mathbf{r},\widetilde{\omega}) + \Delta\omega\frac{\partial\varepsilon}{\partial\omega'}$, we obtain $\frac{\Delta\widetilde{\omega}}{\widetilde{\omega}} = \langle\tilde{\mathbf{e}}^*_\eta|\Delta\varepsilon_w|\mathbf{e}'_{-\eta}\rangle_{V_\text{pert}}$. To remove the dependence of the change on the new perturbed QNM, we then further assume that $\widetilde{\boldsymbol{\Psi}}'_\eta \approx \boldsymbol{\Psi}_\eta$ (first-order Born approximation). Then the denominator amounts to the normalization of Eq. 3 and is equal to one. We obtain

$$\frac{\Delta\widetilde{\omega}}{\widetilde{\omega}} \approx \langle\tilde{\mathbf{e}}^*_\eta|\Delta\varepsilon_w|\mathbf{e}_{-\eta}\rangle_{V_{pert}}, \qquad (5)$$

which predicts the frequency change from the sole knowledge of the initial unperturbed resonance. The impact of the Born approximation on the accuracy is studied in [22] for plasmonic nanoresonators, showing that the cruder approximation is in the numerator and can be refined by local-field corrections. We have not considered these corrections in the tests reported in Fig. 2, obtained for the same grating geometry as in Fig. 1 in another environment $(\varepsilon_w = 1.69)$ instead of air. Figure 2 shows the frequency shifts and linewidth changes predicted with Eq. (5) by a permittivity change of $\varepsilon_w$. Note that the change in $\varepsilon_w$ induces a change in the PML material parameters, which is also considered for computing the integral in Eq. (5), i.e. $V_\text{pert}$ include the modified PML. The predictions are compared with "exact" values computed as the difference of the eigenfrequencies of the perturbed and unperturbed QNMs in two independent computations. Excellent agreement is achieved, especially for small $\Delta\varepsilon_w$.

In conclusion, the present theoretical framework represents a comprehensive contribution to the physics of grating resonances. It allows us to accurately reconstruct the spectral response of gratings in the basis formed by their natural resonances and to analytically and quantitatively predict the shift of their resonance peaks induced by tiny perturbations, directly associating with practical experiments that spectral responses of grating are measured upon plane-wave illumination with fixed incidence angles. A toolbox package, including a $\boldsymbol{\eta}$-QNM solver implemented in the COMSOL Multiphysics environment and Matlab scripts for computing the $\alpha$'s, is available at the webpage of the authors' group [23]. The package is dedicated to one-dimensional gratings for non-conical incidences, and can be straightforwardly generalized to two-dimensional gratings.

## Funding

French National Agency for Research (ANR), project "Resonance" (ANR-16-CE24-0013); French National Agency for Research (ANR) in the frame of "the Investments for the future" Programme IdEx Bordeaux – LAPHIA (ANR-10-IDEX-03-02).

## Acknowledgment

The authors acknowledge the support from the LabEx LAPHIA, the Centre national de la Recherche Scientifique (CNRS), the Agence de l'Innovation de Défense (DGA) and INRIA.